%% file: main.tex
\documentclass[runningheads]{llncs}
\usepackage[T1]{fontenc}
\usepackage{graphicx}
\usepackage{booktabs}
\usepackage{multirow}
\usepackage{color, colortbl}
\usepackage{arydshln}
\usepackage{array}
\usepackage{subcaption}
\usepackage{pifont}
\usepackage{amsmath}
\usepackage{dsfont}
\usepackage{xspace}
\usepackage{hyperref}

\definecolor{Gray}{gray}{0.9}
\definecolor{ashgrey}{rgb}{0.7, 0.75, 0.71}
\definecolor{darkgreen}{rgb}{0,0.39,0}
\newcolumntype{g}{>{\columncolor{Gray}}c}
\newcolumntype{a}{>{\columncolor{ashgrey}}c}

\newcommand{\cmark}{\textcolor{darkgreen}{\ding{51}}}%
\newcommand{\xmark}{\textcolor{red}{\ding{55}}}%

\newcommand{\xmind}{xMIND}
\newcommand{\mind}{MIND}
\newcommand{\polynews}{PolyNews}
\newcommand{\polynewsparallel}{PolyNewsParallel}

\newcommand{\xlt}{\texttt{XLT}\xspace}
\newcommand{\zsxlt}{\texttt{ZS-XLT}}

\newcommand{\xlmrlarge}{XLM-RoBERTa\textsubscript{large}}
\newcommand{\xlmrbase}{XLM-RoBERTa\textsubscript{base}}
\newcommand{\labse}{LaBSE}

\newcommand{\nase}{NaSE}
\newcommand{\nasedae}{NaSE\textsubscript{DAE}}
\newcommand{\nasemt}{NaSE\textsubscript{MT}}
\newcommand{\nasecombined}{NaSE\textsubscript{DAE + MT}}

\newcommand{\tsdae}{TSDAE}

\newcommand{\namlcat}{NAML\textsubscript{CAT}}
\newcommand{\naml}{NAML}
\newcommand{\mins}{MINS}
\newcommand{\caum}{CAUM}
\newcommand{\manner}{MANNeR}

\newcommand{\lfrecce}{LFRec-CE}
\newcommand{\lfrecscl}{LFRec-SCL}

\newcommand{\rparagraph}[1]{\vspace{1.4mm}\noindent\textbf{#1.}}

\newcommand{\sparagraph}[1]{\vspace{0.0mm}\noindent\textbf{#1.}}

\begin{document}
\title{News Without Borders: Domain Adaptation of Multilingual Sentence Embeddings for Cross-lingual News Recommendation}
\titlerunning{News Without Borders}
\author{Andreea Iana\inst{1}\orcidID{0000-0002-7248-7503} \and
Fabian David Schmidt\inst{2}\orcidID{0009-0007-9841-2941} \and
Goran Glavaš\inst{2}\orcidID{0000-0002-1301-6314} \and
Heiko Paulheim\inst{1}\orcidID{0000-0003-4386-8195}}

\authorrunning{A. Iana et al.}
\institute{University of Mannheim, Mannheim, Germany\\
\email{\{andreea.iana, heiko.paulheim\}@uni-mannheim.de}\\
\and
Center For Artificial Intelligence and Data Science, University of Würzburg, Germany
\email{\{fabian.schmidt, goran.glavas\}@uni-wuerzburg.de}}
\maketitle              % typeset the header of the contribution
\begin{abstract}
Rapidly growing numbers of multilingual news consumers pose an increasing challenge to news recommender systems in terms of providing customized recommendations. First, existing neural news recommenders, even when powered by multilingual language models (LMs), suffer substantial performance losses in zero-shot cross-lingual transfer (\zsxlt). Second, the current paradigm of fine-tuning the backbone LM of a neural recommender on task-specific data is computationally expensive and infeasible in few-shot recommendation and cold-start setups, where data are scarce or completely unavailable. In this work, we propose a news-adapted sentence encoder (\nase), domain-specialized from a pretrained massively multilingual sentence encoder (SE). To this end, we compile and leverage \polynews{} and \polynewsparallel{}, two multilingual news-specific corpora. With the news-adapted multilingual SE in place, we test the effectiveness of (i.e., question the need for) supervised fine-tuning for news recommendation, and propose a simple and strong baseline based on (i) frozen \nase{} embeddings and (ii) late click behavior fusion. We show that \nase{} achieves state-of-the-art performance in \zsxlt{} in true cold-start and few-shot news recommendation. 
\end{abstract}

\input{sections/introduction}
\input{sections/related_work}
\input{sections/corpora}
\input{sections/nase}
\input{sections/experimental_setup}
\input{sections/results_discussion}
\input{sections/conclusion}

\begin{credits}
\input{sections/acknowledgements}
\end{credits}

\bibliographystyle{splncs04}
\bibliography{references,anthology}

\begin{appendix}
\input{sections/appendix}
\end{appendix}

\end{document}

%% file: sections/introduction.tex
\section{Introduction}
\label{sec:introduction}

News recommender systems constitute the primary instrument used by digital news platforms to cater to the individual information needs of readers. The ever increasing language diversity of online users has given rise to new challenges for news recommenders. 
Recommender systems must not only produce suitable, balanced, and diverse recommendations for multilingual news consumers from a variety of language backgrounds, but should also accurately perform in cold start scenarios, where news data, user click logs, or both, are missing.

On the one hand, recent advancements in pretrained (multilingual) language models (LMs), used as the backbone of neural news recommenders (NNRs), has allowed to extend NNRs beyond monolingual recommendation \cite{wu2021empowering,guo2023few}. In cross-lingual transfer (\xlt), however, 
even NNRs based on multilingual LMs display drastic performance loss in target-language recommendation compared to their source-language (usually English) recommendation performance \cite{iana2024mind}. Although accurate \xlt{} is critical especially for resource-lean(er) languages with limited-to-no click behavior data, strong multilingual sentence encoders (SEs) -- trained exactly to align sentence semantics across a large number of languages (including many low-resource ones) -- have largely been left unexplored as news encoding backbones in NNRs.

On the other hand, current NNRs typically fine-tune the underlying LM on task-specific data. Not only is fine-tuning a time and resource-intensive process, it is, more critically, infeasible in many real-world scenarios, with: little-to-no news-click data (i) for news in the target language (for particularly low-resource ones) or (ii) for the specific user (the so-called cold-start problem) \cite{wu2024could}. 
%%
% Fine-tuning news encoders based on LMs relies on in-domain data, often not available (e.g., no news and user behavior data for new recommender systems) or expensive and difficult to collect (e.g., news data for low-resource languages). 
Even when using frozen embeddings, most NNRs strictly require in-domain data to learn meaningful user representations for prediction (i.e., to train their parameterized user encoders to aggregate the embeddings of consumed news). 

In this context, multilingual sentence encoders, which align sentence-level semantics across a wide range of languages \cite{yang2019improving,artetxe2019massively,yang-etal-2020-multilingual,feng-etal-2022-language}, hold the promise of reducing the performance gap in \xlt{} for NNRs. 
Multilingual SEs, however, have not been trained for news encoding, i.e., they lack the news domain-specific knowledge, which is crucial for performance in \xlt news recommendation and cold-start scenarios. Moreover, even if equipped with robustly domain-adapted LMs, the majority of NNRs would still require fine-tuning on click behavior data to learn the weights of their trainable user encoder modules.

\vspace{1.4mm}
\noindent\textbf{Contributions.}
We address the above limitations and advance cross-lingual and cold-start news recommendation with the following contributions: 
\textbf{(1)} We compile \polynews{}\footnote{\href{https://huggingface.co/datasets/aiana94/polynews}{https://huggingface.co/datasets/aiana94/polynews}} 
 and \polynewsparallel{}\footnote{\href{https://huggingface.co/datasets/aiana94/polynews-parallel}{https://huggingface.co/datasets/aiana94/polynews-parallel}},
two multilingual news-specific corpora which can be used not only for domain-adaptation of existing LMs, but also for machine translation (MT).
\textbf{(2)} We train a news-adapted multilingual sentence encoder (dubbed \nase) by domain-specializing a general-purpose multilingual SE on \polynews{} and \polynewsparallel{} with denoising auto-encoding and MT objectives.\footnote{\href{https://huggingface.co/aiana94/NaSE}{https://huggingface.co/aiana94/NaSE}}
\textbf{(3)} Leveraging frozen \nase{} embeddings and resorting to non-parameterized late click behavior fusion \cite{iana2023simplifying}, we introduce a simple and strong recommendation technique that yields state-of-the-art performance in zero-shot cross-lingual transfer (\zsxlt) recommendation in cold-start setups, as well as in few-shot recommendation. This challenges the established paradigm of fine-tuning LMs for news recommendation on click behavior data.

%% file: sections/related_work.tex
\section{Related Work}
\label{sec:related_work}

\sparagraph{News Recommendation}
Personalized news recommenders attenuate the information overload for readers by generating suggestions customized to their preferences \cite{li2019survey,wu2023personalized}. Most NNRs comprise a dedicated (i) news encoder, (ii) user encoder, and (iii) click predictor. The news encoder learns news embeddings from various input features \cite{wu2019naml,wu-etal-2019-neural-news,wu2023personalized}, and the user encoder aggregates the embeddings of the users' clicked news into user-level representations \cite{okura2017embedding,an-etal-2019-neural,wu2022news}. Lastly, the recommendation score is computed by comparing the candidate's embedding against the user representation \cite{wang2018dkn,wu2019naml}. NNRs are trained via standard classification objectives \cite{huang2013learning,ijcai2022infonce} and, more recently, contrastive objectives \cite{iana-etal-2024-train,liu2023perconet}.

The existing NNRs have two drawbacks. First, the quality of embeddings produced by news encoders with vanilla multilingual LMs seems inadequate for \xlt, with substantial performance losses for target languages \cite{iana2024mind}. Yet, the usage of multilingual SEs -- precisely trained for cross-lingual semantic alignment across languages on parallel data -- as backbone for news encoders is unexplored. Second, \textit{fine-tuning} of the news encoder (i.e., its LM) is required, which is both (i) computationally expensive, as it updates the LM's hundreds of millions of parameters on large-scale click behavior datasets, and, more critically, (ii) infeasible in setups with limited or no click behavior data (e.g., with news in resource-lean languages or in cold-start setups, with no prior user behavior). In this work, we address these limitations by adapting a general-purpose multilingual SE for the news domain. As the news encoder's backbone, this enables robust \xlt for news recommendation, and supports setups where task-specific fine-tuning is not possible.  

\rparagraph{Domain-Adaptation of Language Models}
The most common approach for injecting domain knowledge into LMs is pretraining on in-domain data with self-supervised objectives, e.g., Masked Language Modeling \cite{devlin-etal-2019-bert} or SimCSE \cite{gao-etal-2021-simcse,liu-etal-2021-fast,wang-etal-2022-english}. Specializing to a particular domain is done either from scratch \cite{beltagy-etal-2019-scibert,wu-etal-2020-tod,lee2020biobert} or by adapting already pretrained LMs \cite{gururangan-etal-2020-dont,hung-etal-2022-ds}. Training from scratch is computationally intensive and requires large-scale domain-specific corpora, often difficult to obtain \cite{wang-etal-2022-gpl,hung-etal-2023-tada}. Adaptation instead is a lighter-weight alternative, as it starts from the already pretrained LM, requiring more moderate amounts of in-domain data. 
For SEs specifically, domain specialization is achieved by training for similarity or relevance estimation tasks with various self-supervised objectives \cite{wang-etal-2022-language,liu2022masked} or via in-domain data generation \cite{wang-etal-2022-gpl}. Current work mainly derives domain-specific SEs from general-purpose (i.e., not sentence-specialized) pretrained LMs. In this work, in contrast, we adapt an existing multilingual SE on in-domain data using denoising auto-encoding and machine translation objectives.

%% file: sections/corpora.tex
\section{Multilingual News Corpora}
\label{sec:corpora}

\begin{figure*}[t]
    \centering
    \includegraphics[width=\textwidth]{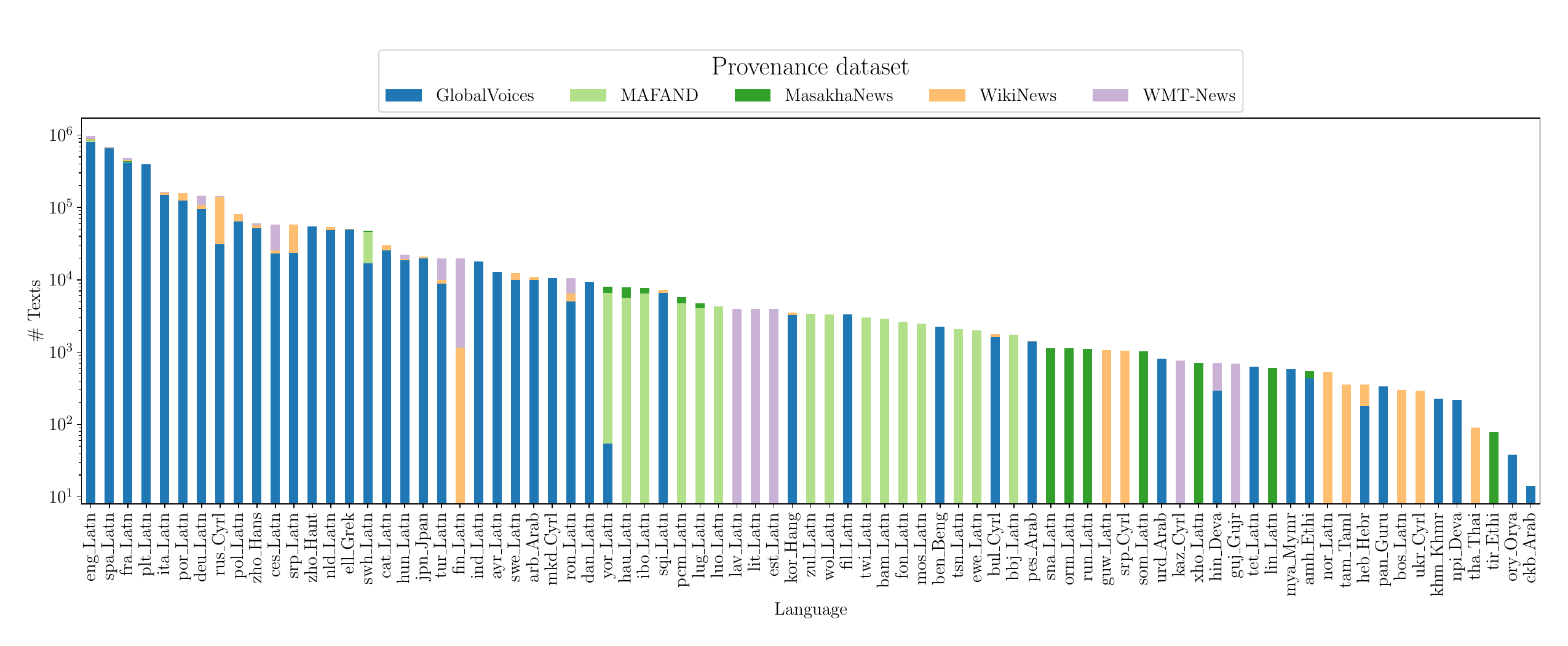}
    \caption{Distribution of texts (log-scale) in \polynews{} across languages and provenance datasets.}
    \label{fig:polynews_language_distrib}
    \vspace{-0.5em}
\end{figure*}

A critical aspect of successful domain adaptation of multilingual LMs is the availability of high-quality training data. To this end, we first compile two large-scale multilingual corpora by leveraging existing news data collections.

\vspace{1.4mm}
\noindent\textbf{PolyNews.}
We compile the multilingual non-parallel corpus \texttt{\polynews{}} by combining news from the following five sources: 
(i) WikiNews (May 2024 dump)\footnote{\href{https://www.wikinews.org/}{https://www.wikinews.org/}}, 
(ii) the \texttt{train} split of MasakhaNews \cite{adelani-etal-2023-masakhanews}, 
(iii) the \texttt{train} split of MAFAND \cite{adelani-etal-2022-thousand}, 
(iv) WMT-News (\texttt{v2019}), 
and (v) GlobalVoices (\texttt{v2018q4}) \cite{tiedemann2012parallel}.
We group together the articles from all five sources according to language-script combinations (some languages, e.g., Serbian, have multiple scripts). 
We preprocess the resulting corpus by removing exact duplicates, and news written in scripts not corresponding to the language of the sub-collection (e.g., Arabic texts in a French or English collection). We use GlotLID \cite{kargaran2023glotlid}, a FastText-based model \cite{joulin2017bag} for language identification. 
In addition, we remove the $K\%$ shortest texts per language (in terms of character length), as these correspond to sequences of words that do not form meaningful sentences.\footnote{We do not use a fixed length, accounting for the varying expressivity of characters in different languages.}
We determine $K$ separately for each of the five news collections, as the quality of the texts heavily depends on their provenance. WikiNews, for example, contains many short (e.g., less than five words) and uninformative texts, which we consider noise.\footnote{We set $K=15\%$ for WikiNews, and $K=3\%$ for the other four sources.}
Lastly, given the importance of avoiding text duplication in LM training \cite{lee-etal-2022-deduplicating}, we perform MinHash \cite{broder1997resemblance} near de-duplication over all sentences of a language.\footnote{We use 256 permutations and an n-gram size of 5. Following manual inspection, we set the de-duplication similarity threshold to 0.9.} 
Finally, \polynews{} contains 3,913,873 news texts, covering 77 languages and 19 scripts. Fig. \ref{fig:polynews_language_distrib} shows its distribution across languages and provenance. We further profile \polynews{} in Appendix \ref{sec:appendix_polynews_lang}.

\vspace{1.4mm}
\noindent\textbf{PolyNewsParallel.}
We compile the translation corpus \texttt{\polynewsparallel{}} from the parallel news collections MAFAND \cite{adelani-etal-2022-thousand}, WMT-News, and GlobalVoices \cite{tiedemann2012parallel}. We use the same preprocessing pipeline as for \polynews{}, and remove near-duplicated texts from the source language. Our final parallel corpus contains 5,386,846 texts over 833 language pairs, spanning 64 languages and 17 scripts.\footnote{Figure \ref{fig:polynewsparallel_language_distrib} shows the distribution of texts across language pairs.}

%% file: sections/nase.tex
\section{News-Adapted Sentence Encoder}
\label{sec:nase}

We obtain the news-adapted sentence encoder \nase{} via sequence-to-sequence training on \polynews{} and \polynewsparallel{} of a massively multilingual sentence encoder. \nase{} can be either (i) further fine-tuned downstream for news recommendation (i.e., on click behavior data) or (ii) directly used as a strong news encoder in cross-lingual news recommendation, without any fine-tuning (see \S\ref{sec:results_discussion}).

\subsection{Domain Adaptation}
\label{sec:domain_adaptation}

Our first training objective (DAE) is an adaption of the transformer-based sequential denoising auto-encoding (\tsdae) approach from \cite{wang2021tsdae}, which we use to specialize a pretrained multilingual SE for the news domain. 
\tsdae{} 
encodes a corrupt version of the input sentence, obtained by adding discrete noise (e.g., token deletion), and then learns to reconstruct the original sentence from the encoding of the noisy input. 
\tsdae{} can be formalized as follows: 

{\footnotesize
    \vspace{-2em}
    \begin{equation}
        \begin{split}
            \mathcal{J}_{TSDAE}(\Theta) &= \mathds{E}_{x \sim D} \left[\log P_\Theta(x | \tilde{x})\right] 
            = \mathds{E}_{x \sim D} \left[\sum_{t=1}^l \log P_\Theta(x | \tilde{x})\right] \\
            &= \mathds{E}_{x \sim D} \left[\sum_{t=1}^l \log \frac{exp(h_t^T e_t)}{\sum_{i=1}^{|V|} exp(h_t^T e_i)} \right]
        \end{split}
    \end{equation}
}
where $D$ is the training corpus, $x = x_1x_2...x_l$ the input sentence with $l$ tokens, $\tilde{x}$ the corresponding corruption, $e_t$ the sequence embedding of $x_t$, $|V|$ the vocabulary size, and $h_t$ the hidden state at decoding step $t$.
At inference, only the encoder is used to produce the embedding for the input text.
We train \nase{} as the \tsdae{} encoder, with the following adjustment. We initialize \nase{} with the pretrained weights of the popular, widely used multilingual SE \labse{} \cite{feng-etal-2022-language}.\footnote{Because we start from an SE rather than vanilla LM, we remove from \tsdae{}'s architecture the layer that pools token-level representation into a sentence embedding.}

Our second training objective is domain-specific sequence-to-sequence machine translation (MT), for which we leverage the parallel data from \polynewsparallel{}. For a given translation pair, we treat the input in the source language to be the `corruption' $\tilde{x}$ of the \textit{target} sentence $x$ in the target language, which is to be `reconstructed'. 
Using these two objectives, DAE and MT, we train four different variants of \nase{} (as the resulting encoder of the corresponding sequence-to-sequence model): 
\textbf{(1) \nasedae} reconstructs the original input sentence from its corruption; 
\textbf{(2) \nasemt} generates the \textit{target}-language translation of the \textit{source}-language input sentence;
\textbf{(3) \nasecombined} combines the two objectives, by randomly choosing either reconstruction or translation for each batch; for parallel data, the reconstruction objective is applied independently on both source- and target-language sentences;
\textbf{(4) \nase\textsubscript{DAE $\rightarrow$ MT}} is trained sequentially, first on reconstruction, and then on translation, i.e., we continue training the \textbf{\nasedae} encoder for translation on parallel data. If not specified otherwise, with just \textbf{\nase{}} we refer to the \nase\textsubscript{DAE $\rightarrow$ MT} variant.
We train \textbf{\nasedae} on \polynews{}, and \textbf{\nasemt} and \textbf{\nasecombined} on \polynewsparallel{}. \textbf{\nase} (i.e., \nase\textsubscript{DAE $\rightarrow$ MT}), as a sequential combination of both training procedures, is trained first on \polynews{}, and then on \polynewsparallel{}.

\subsection{Training Details}
\label{sec:training_details}

\sparagraph{Training Data}
The distributions of both \polynews{} and \polynewsparallel{} are heavily skewed across languages, and the number of news texts for some low-resource languages is particularly very limited.\footnote{E.g., only 100 texts in \polynews{}  for Tigrinya or Kurdish.} We thus follow common practice and smoothen the per-language distribution when sampling for model training \cite{arivazhagan2019massively,conneau-etal-2020-unsupervised,xue-etal-2021-mt5}. We first sample only languages and language pairs that contain at least 100 texts in \polynews{} and \polynewsparallel{}, respectively. We then sample texts from language $L$ by sampling from the modified distribution $p(L) \propto |L|^\alpha$, with $|L|$ as the number of examples in $L$, and $\alpha$ as the smoothing rate hyperparameter ($\alpha < 1$ upsamples 
low-resource, and downsamples high-resource languages). We set $\alpha$ to $0.3$, the value that was found to balance the performance between high- and low-resource languages well \cite{conneau-etal-2020-unsupervised,xue-etal-2021-mt5}.

\rparagraph{Training Settings}
We follow \cite{wang2021tsdae} and use token deletion with a ratio of 0.6 as the discrete corruption for the DAE variants of \nase{}. In all training setups, we tie the encoder and decoder parameters, and initialize them with \labse{} weights. We train each model variant for 50K steps with a learning rate of $3e^{-5}$, using AdamW \cite{loshchilov2018decoupled} as the optimizer. We checkpoint the model every 5K steps. 

\rparagraph{Validation Setup}
We validate \nase{} during training on cross-lingual news recommendation. Concretely, for each model checkpoint, we encode the clicked and candidate news using the \textit{frozen} encoder. We adopt the late fusion approach \cite{iana2023simplifying}, and replace the parameterized user encoder with mean-pooling of dot-product scores between the embeddings of the candidate and the clicked news. 
Eliminating a parameterized user encoder has two benefits: (1) it increases the computational efficiency of \nase{} validation, as we do not have to train the NNR (i.e., its user encoder), and (2) it isolates \nase{} (i.e., the news encoder), as the only component that affects the recommendation performance, eliminating the user encoder as a confounding factor.

\rparagraph{Validation Data}
A robust multilingual SE should produce good embeddings in many languages. We thus validate the quality of the \nase{} embeddings during training---for the purposes of model (i.e., optimal checkpoint) selection---on  the \textit{small} variants of the English \mind{} \cite{wu-etal-2020-mind}, and its multilingual \xmind{} \cite{iana2024mind} counterpart. \xmind{} comprises news articles from \mind{} machine-translated into 14 languages. To measure recommendation performance, we combine the multilingual news articles from \xmind{} with the user click behavior data from \mind{}; our final validation set covers user behaviors from the last day of the \mind{} training set. Our resulting validation corpus covers 15 linguistically diverse languages, offering a more realistic estimate of the quality of multilingual sentence embeddings produced by \nase{} for \xlt in news recommendation.

%% file: sections/experimental_setup.tex
\section{Experimental Setup}
\label{sec:experimental_setup}

Zero-shot cross-lingual (\zsxlt{}) news recommendation is the task for which we primarily develop \nase{}, and thus, the downstream task on which we evaluate it. In all experiments, we assume only monolingual news consumption, i.e., that each user reads news only in one language and, accordingly, also receives recommendations in one language.

\rparagraph{Neural News Recommenders}
We evaluate four architecturally diverse NNRs which showed promising results on the \xmind{} dataset \cite{iana2024mind}: (1)\textit{ \naml{}} \cite{wu2019naml}, (2) \textit{\mins{}} \cite{wang2022news}, (3) \textit{\caum{}} \cite{qi2022news}, and (4) \textit{\manner{}} \cite{iana-etal-2024-train}.\footnote{Only the base version with the \texttt{CR-Module}, without any \texttt{A-Module} for aspect-based diversification.}
Additionally, we consider three simpler yet competitive baselines: (5) \textit{\lfrecce{}}, (6) \textit{\lfrecscl{}}, and (7) \textit{\namlcat{}}. 
\naml{}, \mins{}, and \caum{} were designed to encode textual information using pretrained word embeddings \cite{pennington2014glove}, contextualized with convolution neural networks (CNNs) \cite{kim-2014-convolutional} or attention layers \cite{bahdanau2014neural}, whereas \lfrecce{}, \lfrecscl{}, and \manner{} use a pretrained LM as their news encoder. In addition to the news text, \naml{}, \mins{}, and \caum{} leverage category information, while \manner{} and \caum{} encode named entities extracted from the title and abstract of the news. %
Lastly, \namlcat{} is a text-agnostic variant of \naml{} that learns news embeddings purely as randomly initialized and fine-tuned category vectors.
Following \cite{wu2021empowering}, we replace the original news encoders of NAML, MINS, and CAUM with a pretrained LM in order to enable multilingual recommendation, and to ensure a fair comparison between models.\footnote{\naml, \mins, and \caum{} pool token embeddings with an attention layer to obtain the sentence embedding from the LM, whereas \manner{} uses the vector of the \texttt{CLS} token.} Concretely, we experiment with (1) the pretrained multilingual LM \xlmrlarge \cite{conneau-etal-2020-unsupervised}, (2) the BERT-based multilingual SE \labse{} \cite{feng-etal-2022-language}, and (3) our proposed \nase{}. 

The recommenders also differ in the way they model users. Specifically, \naml{} \cite{wu2019naml} encodes their preferences using additive attention \cite{bahdanau2014neural}, while \mins{} \cite{wang2022news} combines multi-head self-attention \cite{vaswani2017attention} with a multi-channel GRU-based recurrent network \cite{cho2014learning} and additive attention. The more complex CAUM \cite{qi2022news} uses a candidate-aware self-attention network to learn long-term user preferences, and a candidate-aware CNN to capture the users' short-term interests. It obtains the final user embeddings by attending over the two intermediate representations. In contrast to these models, \manner{} \cite{iana-etal-2024-train}, \lfrecce{} and \lfrecscl{} resort to the non-parameterized late fusion approach of \cite{iana2023simplifying} for aggregating click behaviors.
With the exception of \manner{} and \lfrecscl{}, which minimize the supervised contrastive loss (SCL) \cite{khosla2020supervised}, the remaining recommenders are trained by minimizing the standard cross-entropy (CE) loss.

\input{tables/mind_stats}

\vspace{1.4mm}
\noindent\textbf{Data.}
We conduct experiments on the \textit{small} variants of the English \mind{} \cite{wu-etal-2020-mind} and the multilingual \xmind{} \cite{iana2024mind} datasets. As mentioned in \S\ref{sec:training_details}, we couple the news texts from \xmind{} with the click behavior data from \mind{}, via news IDs. Wu et al. \cite{wu-etal-2020-mind} do not release test labels for \mind{}. Hence, we use the validation set for testing, and split the training set into temporarily disjoint portions for training (first four days) and validation (last day), as per Table \ref{tab:mind_stats}.

\vspace{1.4mm}
\noindent\textbf{Fine-Tuning Details.}
In all experiments that require LM fine-tuning, we update only the LM's last four layers.\footnote{In early experiments with \xlmrbase{}, fine-tuning the whole LM did not bring gains compared to updating only the last four layers. For computational efficiency, we thus keep the bottom eight layers of \labse{} and \nase{}, and the bottom 20 layers of \xlmrlarge{}, frozen.}
We set the maximum click history length to 50, and sample four negatives per positive in training, as per \cite{ijcai2022infonce}. 
We tune the main hyperparameters of all NNRs using grid search. More specifically, we search for the optimal learning rate in $\{1e^{-3}, 1^{e-4}, 1^{e-5}\}$, finding $1e^{-5}$ to be the most suitable value for all NNRs. We optimize the number of heads in the multi-head self-attention networks of \naml, \mins, \caum{} in $[8, 12, 16, 24, 32]$, and the query vector dimensionality in the additive attention network of \naml{} and \mins{} in $[50, 200]$, with a step of 50. We use the following best performing settings: 32 attention heads for \naml{} and \mins{}, 8 for \caum{} when coupled with \xlmrlarge, and 24, 16, and 8 attention heads for \naml{}, \mins{}, and \caum{}, respectively, when using \labse{} or \nase{} as the news encoder's backbone. Moreover, we use a query vector of dimension 200 for \naml{} and 50 for \mins{} in combination with \xlmrlarge, whereas for SE-based news encoders we use a dimensionality of 50 for \naml{} and 100 for \mins{}. We find the optimal temperature of 0.38 for the supervised contrastive loss in \manner{} and \lfrecscl{} by sweeping the interval $[0.1, 0.5]$, with a step of 0.02. We set all the remaining model-specific hyperparameters to the optimal values reported in the respective papers. 
We train the models for 10 epochs, with a batch size of 8 for the SE-based variants, and 4 for the \xlmrlarge{}-equipped NNRs. We train using mixed precision, and the Adam optimizer \cite{kingma2014adam}. 
We repeat each experiment three times with the seeds $\{42, 43, 44\}$, set with PyTorch Lightning's \texttt{seed\_everything}, and report the mean and standard deviations for common metrics: AUC, MRR, and nDCG@10.\footnote{For brevity, we omit results for nDCG@5, as they exhibit the same patterns as nDCG@10.}
We train \nase{} using Sentence Transformers \cite{reimers-gurevych-2019-sentence} and PyTorch Lightning \cite{Falcon_PyTorch_Lightning_2019} on a cluster with virtual machines, on single NVIDIA A100 40GB GPUs.\footnote{Code available at \href{https://github.com/andreeaiana/nase}{https://github.com/andreeaiana/nase}. The implementation also provides the steps to create the \polynews{} and \polynewsparallel{} datasets.}
We conduct all news recommendation experiments with the NewsRecLib library \cite{iana-etal-2023-newsreclib}, on the same cluster, on single NVIDIA A40 48GB GPUs.

%% file: tables/mind_stats.tex
\begin{table}[t]
    \centering
    
    \caption{\mind{} (\textit{small}) and \xmind{} (\textit{small}) statistics. Note that for \xmind{} we report the statistics \textit{per language}, i.e., in total \xmind{} contains 14 languages.}
    \label{tab:mind_stats}
    
    \begin{tabular}{llcc}
        \toprule
        & \textbf{Train} & \textbf{Validation} & \textbf{Test} \\ \midrule

        \# News & 51,282 & 51,282 & 42,416 \\
        \# Users & 45,214 & 19,703 & 48,593 \\
        \# Impressions & 124,229 & 29,498 & 70,938 \\
        
        \bottomrule
    \end{tabular}
    
    \vspace{-0.5em}
\end{table}

%% file: sections/results_discussion.tex
\section{Results and Discussion}
\label{sec:results_discussion}

We compare the NNRs' performance for news encoders with different LM/SE backbones in cross-lingual transfer for news recommendation. In this setup, the user history and candidates during training are \textit{monolingual} and in \textit{English} only. At inference, both the user history and the candidate news are solely in one of the 14 \textit{target languages} of \xmind{}. 

\rparagraph{Fine-Tuning News Encoders}
We first investigate the standard recommendation, in which task-specific data (i.e., click behavior information and news impressions) are used to train the NNR. Table \ref{tab:zsxlt_rec_finetuned_lm} displays the performance of the recommenders when the underlying news encoders are fine-tuned (i.e., the weights of the backbone LM or SE are updated) for news recommendation, on English data. 
%%%

While SE-powered NNRs outperform the text-agnostic baseline \namlcat{}, the NNRs relying on \xlmrlarge{} sometimes underperform this simple baseline. 
Similarly, \xlmrlarge{} underperforms \labse{} and \nase{} on both English recommendation and in \zsxlt{}, regardless of the NNR in which it is used. We believe this is because the \xlmrlarge{}-based encoder first needs to learn how to aggregate token representations into sentence embeddings. These results clearly render (multilingual) SEs, pretrained to produce robust sentence embeddings, beneficial for news recommendation.
%%%
SE-based NNRs achieve similar performance with \labse{} and \nase{}, both in English, and in \zsxlt{} on  \xmind{}, i.e., \nase{} does not bring gains over \labse{} from which we derived it.
% \footnote{Appendix \ref{sec:appendix_lang_zsxlt} provides detailed per-language results.}
%
We argue that this is due to the fact that fine-tuning on news recommendation also leads to domain adaptation: \labse{} itself becomes sufficiently specialized for the news domain through large-scale recommendation fine-tuning, compensating for \nase{}'s task-agnostic domain adaptation.

Lastly, we note that the simple baselines \lfrecce{}, and in particular \lfrecscl{}, exhibit strong recommendation performance, often surpassing more complex models like \caum{} or \mins{}, which have richer input (e.g., topical categories, named entities) and parameterized user encoders. This is in line with the findings of Iana et al. \cite{iana2023simplifying} and questions the need for complex parameterized user encoders. 
\input{tables/zsxlt_rec_finetuned_lm}
\input{tables/zsxlt_rec_frozen_lm}

\rparagraph{Frozen News Encoders}
Fine-tuning news encoder backbones (LMs with hundreds of millions of parameters) on large-scale recommendation data can be prohibitively expensive for many practitioners without access to large computational resources (e.g., GPUs). We thus next analyze how NNRs perform with frozen news encoders (i.e., no updates to LM/SE), allowing updates only to other (fewer) trainable parameters of the news recommenders. 
Specifically, the model can now only learn how to encode other input features (e.g., categories), or how to aggregate the news embeddings into a user-level encoding, if equipped with parameterized user encoders. For most models, freezing the news encoder reduces the number of trainable parameters by two orders of magnitude.

Table \ref{tab:zsxlt_rec_frozen_lm} summarizes the recommendation performance of the NNRs with frozen news encoders. Unsurprisingly, \xlmrlarge{}-based recommenders yield the weakest performance across all languages, as the LM itself cannot be tuned to encode token sequences. Out-of-the-box \nase{} embeddings substantially outperform \labse{} in most cases cases: e.g., for nDCG@10 by 2.58\% on English, and 4.17\% cross-lingually (averaged over all 14 \xmind{} languages), averaged across all NNRs. 
The performance gap between \labse{} and \nase{} becomes smaller when the NNR uses a more complex \textit{trainable} user encoder (e.g., \mins{}): the user encoder parameters take over the domain-specialization task in large-scale fine-tuning.   

Two key aspects point to successful (task-agnostic) domain-specialization of \nase{}. \textbf{1)} We observe that for models without trainable user encoders the relative performance loss in \zsxlt{} compared to English performance is less pronounced with \nase{} as news encoder than with \labse{}. As target-language data is often scarce in many real-world applications, closing the gap to performance on the source language on which the recommender is trained is crucial for multilingual news recommendation. 
\textbf{2)} Unlike \xlmrlarge{} or \labse{}, who both benefit from large-scale recommendation fine-tuning, \nase{}'s performance when frozen is much closer to its fine-tuned performance. This calls into question the current paradigm of performing expensive supervised fine-tuning of the NNR's news encoder. 

Most importantly, coupled with a frozen \nase{} NE, \lfrecscl{} -- a model with no other trainable parameters -- achieves state-of-the-art performance over more complex, trainable NNRs. On the one hand, this demonstrates the effectiveness of the news-specialized \nase{} encoder (trained in a task-agnostic manner) over a general-purpose SE like \labse{} in news recommendation. 
%%%
On the other hand, it proves that \lfrecscl{} can produce good recommendations in true cold-start setup, where no news or user data is available (for training a parameterized NNR). This is in contrast to all the other models that require task-specific data to learn meaningful user representations, which are then to compute the final recommendation scores. 
%%%
Notably, this means that with \lfrecscl{} with (frozen) \nase{} we obtain a state-of-the-art news recommendation performance without the need for any task-specific training for news recommendation. \lfrecscl{} with (frozen) \nase{} outperforms more complex models (\naml{}, \mins{}, \caum{}), with user encoders trained for recommendation, in terms of ranking (i.e., MRR, nDCG@$k$), and performs on-par or better in terms of classification (i.e., AUC). These results suggest that domain-specialization of a multilingual SE (i.e., \nase{}) removes the need for supervised training of NNRs.   

\begin{figure}[t]
    \centering
    \includegraphics[width=\textwidth]{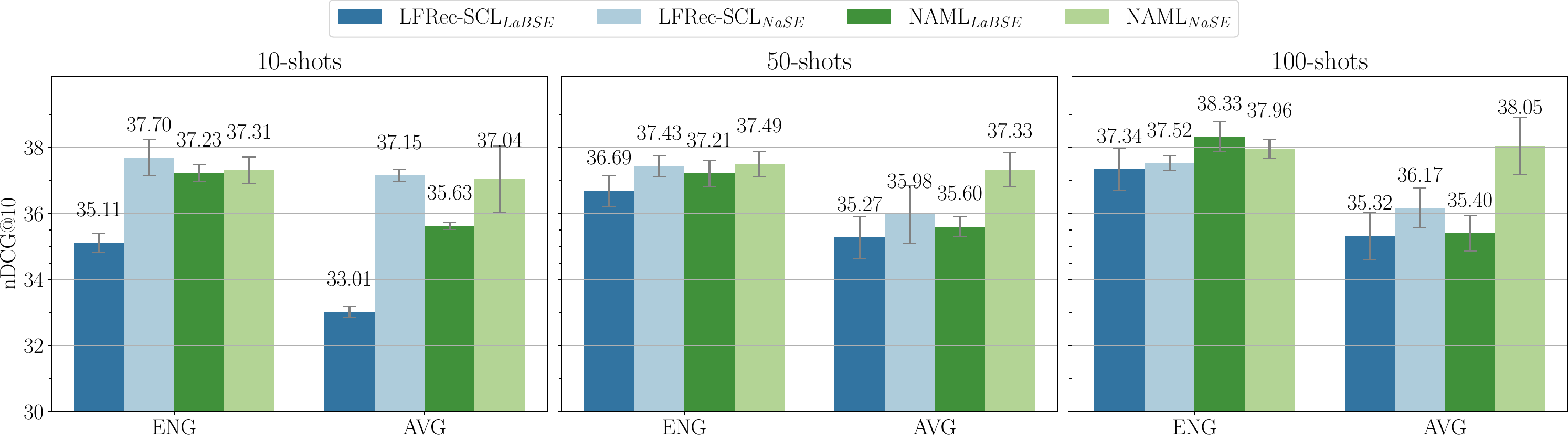}
    \caption{\zsxlt{} performance in few-shot recommendation (i.e., for different number of impressions used in training). For each model, we report average performance across three runs (i) on the English MIND dataset (denoted \texttt{ENG}), and (ii) averaged across all 14 target languages of xMIND (denoted \texttt{AVG}).}
    \label{fig:fs_rec}
    \vspace{-0.5em}
\end{figure}

\vspace{1.4mm}
\noindent\textbf{Few-Shot Recommendation.}
To corroborate the previous findings, we also investigate the performance of the NNRs when the underlying news encoder is fine-tuned on just a few task-specific examples, namely in \textit{few-shot recommendation}. Here, we assume only a handful of task-specific examples (i.e., impressions) for training the NNRs. In terms of \xlt, we stick to zero-shot \xlt for recommendation, i.e., we assume that the few training instances that we have are all in the source language, i.e., English.\footnote{This should not be confused with few-shot \xlt for recommendation, which assumes (few) target-language instances in the user histories during training, but overall relies on large amounts of task-specific training data  \cite{iana2024mind}.}
Fig. \ref{fig:fs_rec} shows the results for 10, 50, and 100 shots, i.e., the number of impressions used to train the NNR.
% \footnote{Appendix \ref{sec:appendix_lang_zsxlt} provides detailed per-language results.} 
We again observe that, when fine-tuned on small amounts of data, \nase{} as the news encoder generally outperforms \labse{}, especially for fewer shots (10 and 50). In 10-shot recommendation, \nase{}'s relative gain in terms of nDCG@10 over \labse{} ranges from 3.96\% (averaged over 14 target languages) and 0.21\% on English (for \naml) to 12.54\%  and 7.38\% (for \lfrecscl), respectively. The differences between the two decrease with more training data (100 shots). Besides cold starts, these results render \nase{} effective in \zsxlt{} recommendation in realistic low-data setups.

\begin{figure}[t]
    \centering
    \includegraphics[width=\textwidth]{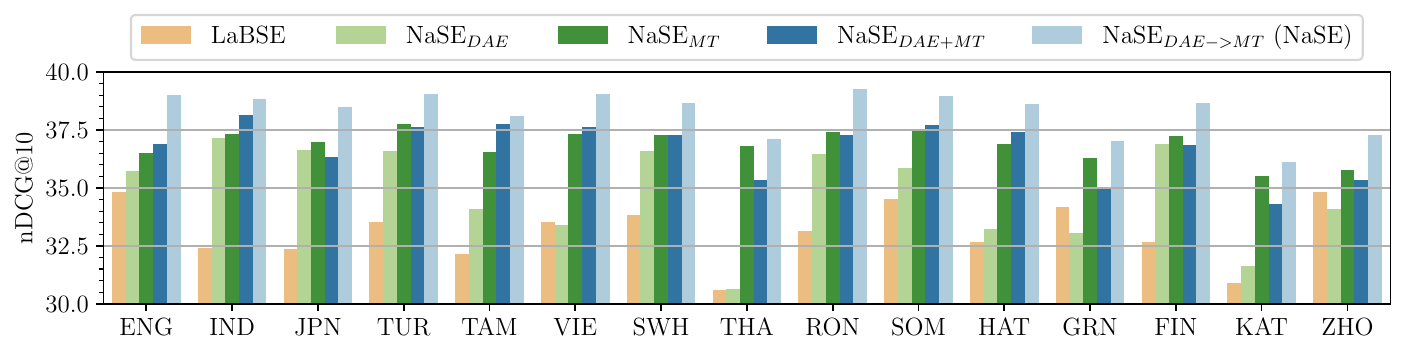}
    \caption{Ranking performance (nDCG@10) of \lfrecscl{}{} with frozen \nase{} embeddings for different training strategies of \nase{} versus \lfrecscl{} with frozen \labse{} embeddings, over English (\texttt{ENG}) and the 14 languages of \xmind.}
    \label{fig:ablation_training_ndcg}
    \vspace{-0.5em}
\end{figure}

\rparagraph{Impact of Training Strategy}
Lastly, we ablate \nase's performance for different training strategies. Fig. \ref{fig:ablation_training_ndcg} shows the ranking performance (i.e., nDCG@10) of \lfrecscl{} with frozen \nase{} embeddings.
We find that the denoising auto-encoding  pretraining objective (\texttt{DAE}) underperforms all other \nase{} configurations. Nonetheless, even \lfrecscl{} with frozen \nasedae{} embeddings achieves between 0.19\% (on \texttt{THA}) and 12.90\% (on \texttt{FIN}) relative improvements over the model using frozen \labse{} embeddings, outperforming it on 12 out of 15 languages.
The translation objective seems highly effective for all languages: 
\nasemt{} brings up to 19.99\% and 20.22\% improvements relative to \nasedae{}, and \labse{}, respectively. Interestingly, the gains seem largest for languages not present in \polynewsparallel{} on which we train \nase{} (e.g., \texttt{THA}) and those not seen by \labse{} in pretraining (e.g., \texttt{GRN}): this hints at positive cross-lingual transfer of domain-specialization via the MT training.

Training on both \texttt{DAE} and \texttt{MT} slightly improves \texttt{DAE}-only specialization for most languages. However, \textit{sequentially} combining the two (\nase\textsubscript{DAE $\rightarrow$ MT}; referred to with just \nase{}) brings further gains and overall best performance. We believe that this is due to the sequential training being exposed to more data, i.e., both \polynews{} (in the \texttt{DAE} stage) and \polynewsparallel{} (in the \texttt{MT} stage).
Compared to \labse{}, \nase{} in this setup yields relative improvements ranging from 7.05\% on \texttt{ZHO}, to 11.99\% on \texttt{ENG}, and 21.33\% on \texttt{THA}.

It is worth emphasizing that news-specialization of \nase{} with a translation-based objective benefits also languages \textit{not} included in \polynewsparallel{} (e.g., \texttt{THA}, \texttt{HAT}, \texttt{KAT}), and even languages not present in \labse{}'s large-scale parallel pretraining corpus (e.g., \texttt{GRN}). This points to the language-agnostic nature of news-specific domain specialization. This finding is in line with results from prior work in machine translation \cite{tran2020cross,duquenne2023sonar}, which showed that modest amounts of parallel data (as in our \polynewsparallel{}) improve the cross-lingual semantic alignment in multilingually trained encoder-decoder models. This then results in performance gains even for languages not present in the parallel corpora.

%% file: tables/zsxlt_rec_finetuned_lm.tex
\begin{table*}[t]
    \centering

    \caption{\zsxlt{} recommendation performance with a fine-tuned news encoder (LM/SE). For each model, we report its size in terms of millions of trainable and total parameters, and performance (i) on the English MIND (\texttt{ENG}), (ii) averaged across all 14 target languages of \xmind{} (\texttt{AVG}), and (iii) the relative percentage difference between average \zsxlt{} and \texttt{ENG} performance ($\%\Delta$). Reported performance is averaged over three runs. Subscripts denote standard deviation.}
    \label{tab:zsxlt_rec_finetuned_lm}
    
    \resizebox{\textwidth}{!}{%
    \begin{tabular}{lllcgacgacga}
        \toprule
        \multirow{2}{*}{\textbf{Model}} & \multicolumn{2}{c}{\textbf{\# Parameters (M)}}
        & \multicolumn{3}{c}{\textbf{AUC}} & \multicolumn{3}{c}{\textbf{MRR}} 
        % & \multicolumn{3}{c}{\textbf{nDCG@5}} 
        & \multicolumn{3}{c}{\textbf{nDCG@10}} 
        \\ \cmidrule(lr){2-3} \cmidrule(lr){4-6} \cmidrule(lr){7-9} \cmidrule(lr){10-12} 
        & Trainable & Total
        & ENG & AVG & $\%\Delta$ 
        & ENG & AVG & $\%\Delta$ 
        % & ENG & AVG & $\%\Delta$ 
        & ENG & AVG & $\%\Delta$ \\ \hline

        \naml\textsubscript{CAT}
        & 0.387
        & 0.387
        & \multicolumn{2}{c}{55.46\textsubscript{$\pm$0.18}}  % AUC 
        & 0.00
        & \multicolumn{2}{c}{31.12\textsubscript{$\pm$0.56}}  % MRR 
        & 0.00
        % & \multicolumn{2}{c}{29.44\textsubscript{$\pm$0.67}}  % nDCG@5 
        % & 0.00
        & \multicolumn{2}{c}{35.81\textsubscript{$\pm$0.59}}  % nDCG@10 
        & 0.00
        \\ \hdashline

        \lfrecce\textsubscript{\xlmrlarge}
        & 307
        & 559
        & 50.00\textsubscript{$\pm$0.00} % AUC ENG
        & 50.00\textsubscript{$\pm$0.00} % AUC AVG
        & 0.00 % Delta
        & 35.71\textsubscript{$\pm$0.70}  % MRR ENG
        & 33.05\textsubscript{$\pm$0.60}  % MRR AVG
        & -7.45 % Delta
        % & 33.97\textsubscript{$\pm$0.87}  % nDCG@5 ENG
        % & 31.28\textsubscript{$\pm$0.70} % nDCG@5 AVG
        % & -7.92 % Delta
        & 40.32\textsubscript{$\pm$0.77}  % nDCG@10 ENG
        & 37.66\textsubscript{$\pm$0.65}  % nDCG@10 AVG
        & -6.59 % Delta
        \\

        \lfrecce\textsubscript{\labse}
        & 414
        & 470
        & 62.89\textsubscript{$\pm$1.50} % AUC ENG
        & 62.20\textsubscript{$\pm$1.08} % AUC AVG
        & -1.10% Delta
        & 36.71\textsubscript{$\pm$0.53}  % MRR ENG
        & 35.65\textsubscript{$\pm$0.50} % MRR AVG
        & -2.90 % Delta
        % & 35.04\textsubscript{$\pm$0.67}  % nDCG@5 ENG
        % & 33.88\textsubscript{$\pm$0.62} % nDCG@5 AVG
        % & -3.30 % Delta
        & 41.30\textsubscript{$\pm$0.60}  % nDCG@10 ENG
        & 40.15\textsubscript{$\pm$0.56}   % nDCG@10 AVG
        & -2.77 % Delta
        \\

        \lfrecce\textsubscript{\nase}
        & 414
        & 470
        & 65.19\textsubscript{$\pm$0.23} % AUC ENG
        & 63.00\textsubscript{$\pm$1.71} % AUC AVG
        & -3.35 % Delta
        & 36.24\textsubscript{$\pm$0.47}  % MRR ENG
        & 33.39\textsubscript{$\pm$1.07}  % MRR AVG
        & -7.85 % Delta
        % & 34.31\textsubscript{$\pm$0.48}  % nDCG@5 ENG
        % & 31.54\textsubscript{$\pm$1.13} % nDCG@5 AVG
        % & -8.07 % Delta
        & 40.74\textsubscript{$\pm$0.44}  % nDCG@10 ENG
        & 37.92\textsubscript{$\pm$1.13}  % nDCG@10 AVG
        & -6.93 % Delta
        \\ \hdashline

        \lfrecscl\textsubscript{\xlmrlarge}
        & 307
        & 559
        & 50.00\textsubscript{$\pm$0.00} % AUC ENG
        & 50.00\textsubscript{$\pm$0.00} % AUC AVG
        & 0.00 % Delta
        & 29.12\textsubscript{$\pm$0.77}  % MRR ENG
        & 27.88\textsubscript{$\pm$0.84}  % MRR AVG
        & -4.27 % Delta
        % & 26.89\textsubscript{$\pm$0.71}  % nDCG@5 ENG
        % & 25.61\textsubscript{$\pm$0.91} % nDCG@5 AVG
        % & -4.76 % Delta
        & 33.28\textsubscript{$\pm$0.81}  % nDCG@10 ENG
        & 31.96\textsubscript{$\pm$0.90}  % nDCG@10 AVG
        & -3.96 % Delta
        \\

        \lfrecscl\textsubscript{\labse}
        & 414
        & 470
        & 68.94\textsubscript{$\pm$0.87} % AUC ENG
        & 67.29\textsubscript{$\pm$0.75} % AUC AVG
        & -2.39% Delta
        & 36.45\textsubscript{$\pm$0.65} % MRR ENG
        & 35.04\textsubscript{$\pm$0.53}  % MRR AVG
        & -3.87 % Delta
        % & 34.93\textsubscript{$\pm$0.62}  % nDCG@5 ENG
        % & 33.42\textsubscript{$\pm$0.49} % nDCG@5 AVG
        % & -4.32 % Delta
        & 41.14\textsubscript{$\pm$0.60}  % nDCG@10 ENG
        & 39.65\textsubscript{$\pm$0.47}  % nDCG@10 AVG
        & -3.62 % Delta
        \\

        \lfrecscl\textsubscript{\nase}
        & 414
        & 470
        & 68.63\textsubscript{$\pm$1.23} % AUC ENG
        & 66.73\textsubscript{$\pm$0.91} % AUC AVG
        & -2.76% Delta
        & 36.59\textsubscript{$\pm$0.87}  % MRR ENG
        & 34.78\textsubscript{$\pm$0.61}  % MRR AVG
        & -4.95 % Delta
        % & 34.80\textsubscript{$\pm$0.87} % nDCG@5 ENG
        % & 33.09\textsubscript{$\pm$0.57} % nDCG@5 AVG
        % & -4.93 % Delta
        & 41.07\textsubscript{$\pm$0.82}  % nDCG@10 ENG
        & 39.35\textsubscript{$\pm$0.52}  % nDCG@10 AVG
        & -4.17 % Delta
        \\ \hdashline
       
        \naml\textsubscript{\xlmrlarge}
        & 312
        & 564
        & 55.05\textsubscript{$\pm$4.35} % AUC ENG
        & 53.54\textsubscript{$\pm$3.28} % AUC AVG
        & -2.72 % Delta
        & 34.36\textsubscript{$\pm$1.28}  % MRR ENG
        & 32.60\textsubscript{$\pm$0.69}  % MRR AVG
        & -5.12 % Delta
        % & 32.57\textsubscript{$\pm$1.50}  % nDCG@5 ENG
        % & 30.74\textsubscript{$\pm$0.85} % nDCG@5 AVG
        % & -5.64 % Delta
        & 38.90\textsubscript{$\pm$1.40}  % nDCG@10 ENG
        & 37.04\textsubscript{$\pm$0.90}  % nDCG@10 AVG
        & -4.77 % Delta
        \\

        \naml\textsubscript{\labse}
        & 414
        & 471
        & 51.59\textsubscript{$\pm$0.71} % AUC ENG
        & 51.93\textsubscript{$\pm$0.65} % AUC AVG
        & +0.66 % Delta
        & 36.63\textsubscript{$\pm$0.68}  % MRR ENG
        & 35.91\textsubscript{$\pm$0.51}  % MRR AVG
        & -1.98 % Delta
        % & 34.93\textsubscript{$\pm$0.80}  % nDCG@5 ENG
        % & 34.14\textsubscript{$\pm$0.53} % nDCG@5 AVG
        % & -2.27 % Delta
        & 41.25\textsubscript{$\pm$0.71}  % nDCG@10 ENG
        & 40.42\textsubscript{$\pm$0.49}  % nDCG@10 AVG
        & -1.99 % Delta
        \\

        \naml\textsubscript{\nase}
        & 414
        & 471
        & 53.65\textsubscript{$\pm$0.90} % AUC ENG
        & 51.06\textsubscript{$\pm$0.54} % AUC AVG
        & -4.83 % Delta
        & 36.97\textsubscript{$\pm$0.68}  % MRR ENG
        & 34.38\textsubscript{$\pm$0.95}  % MRR AVG
        & -7.00 % Delta
        % & 35.24\textsubscript{$\pm$0.74}  % nDCG@5 ENG
        % & 32.61\textsubscript{$\pm$0.98} % nDCG@5 AVG
        % & -7.46 % Delta
        & 41.62\textsubscript{$\pm$0.66}  % nDCG@10 ENG
        & 38.94\textsubscript{$\pm$0.93}  % nDCG@10 AVG
        & -6.43 % Delta
        \\
        \hdashline

        \mins\textsubscript{\xlmrlarge}
        & 316
        & 568
        & 52.18\textsubscript{$\pm$2.86} % AUC ENG
        & 53.43\textsubscript{$\pm$3.00} % AUC AVG
        & +2.40 % Delta
        & 31.64\textsubscript{$\pm$1.98}  % MRR ENG
        & 30.95\textsubscript{$\pm$2.19}  % MRR AVG
        & -2.19 % Delta
        % & 29.74\textsubscript{$\pm$2.13}  % nDCG@5 ENG
        % & 29.00\textsubscript{$\pm$2.34} % nDCG@5 AVG
        % & -2.49 % Delta
        & 36.12\textsubscript{$\pm$2.05}  % nDCG@10 ENG
        & 35.39\textsubscript{$\pm$2.24}  % nDCG@10 AVG
        & -4.77 % Delta
        \\

        \mins\textsubscript{\labse}
        & 416
        & 473
        & 63.70\textsubscript{$\pm$1.02} % AUC ENG
        & 62.31\textsubscript{$\pm$0.77} % AUC AVG
        & -2.19 % Delta
        & 34.31\textsubscript{$\pm$0.80}  % MRR ENG
        & 33.58\textsubscript{$\pm$0.58} % MRR AVG
        & -2.11 % Delta
        % & 32.52\textsubscript{$\pm$0.80}  % nDCG@5 ENG
        % & 31.72\textsubscript{$\pm$0.54} % nDCG@5 AVG
        % & -2.46 % Delta
        & 39.06\textsubscript{$\pm$0.69}  % nDCG@10 ENG
        & 38.22\textsubscript{$\pm$0.46}  % nDCG@10 AVG
        & -2.17 % Delta
        \\

        \mins\textsubscript{\nase}
        & 416
        & 473
        & 62.33\textsubscript{$\pm$0.95} % AUC ENG
        & 60.81\textsubscript{$\pm$1.23} % AUC AVG
        & -2.44 % Delta
        & 34.54\textsubscript{$\pm$0.89}  % MRR ENG
        & 33.53\textsubscript{$\pm$0.67}  % MRR AVG
        & -2.92 % Delta
        % & 32.79\textsubscript{$\pm$1.06}  % nDCG@5 ENG
        % & 31.83\textsubscript{$\pm$0.73} % nDCG@5 AVG
        % & -2.92 % Delta
        & 39.36\textsubscript{$\pm$0.86}  % nDCG@10 ENG
        & 38.35\textsubscript{$\pm$0.62} % nDCG@10 AVG
        & -2.58 % Delta
        \\
        \hdashline

        \caum\textsubscript{\xlmrlarge}
        & 316
        & 568
        & 58.67\textsubscript{$\pm$0.41} % AUC ENG
        & 59.11\textsubscript{$\pm$0.74}  % AUC AVG
        & +0.74% Delta
        & 30.01\textsubscript{$\pm$2.30}  % MRR ENG
        & 30.04\textsubscript{$\pm$1.98}  % MRR AVG
        & +0.08 % Delta
        % & 28.35\textsubscript{$\pm$2.50}  % nDCG@5 ENG
        % & 28.37\textsubscript{$\pm$2.26} % nDCG@5 AVG
        % & +0.06 % Delta
        & 34.57\textsubscript{$\pm$2.29}  % nDCG@10 ENG
        & 34.57\textsubscript{$\pm$2.07}  % nDCG@10 AVG
        & 0.00 % Delta
        \\

        \caum\textsubscript{\labse}
        & 417
        & 474
        & 64.92\textsubscript{$\pm$0.83} % AUC ENG
        & 63.52\textsubscript{$\pm$0.80} % AUC AVG
        & -2.16% Delta
        & 34.36\textsubscript{$\pm$0.40}  % MRR ENG
        & 33.53\textsubscript{$\pm$0.69} % MRR AVG
        & -2.69 % Delta
        % & 32.82\textsubscript{$\pm$0.19}  % nDCG@5 ENG
        % & 31.73\textsubscript{$\pm$0.55} % nDCG@5 AVG
        % & -3.33 % Delta
        & 39.40\textsubscript{$\pm$0.21}  % nDCG@10 ENG
        & 38.34\textsubscript{$\pm$0.54}  % nDCG@10 AVG
        & -2.70 % Delta
        \\

        \caum\textsubscript{\nase}
        & 417
        & 474
        & 64.54\textsubscript{$\pm$0.53} % AUC ENG
        & 62.54\textsubscript{$\pm$0.37} % AUC AVG
        & -3.10 % Delta
        & 35.24\textsubscript{$\pm$0.40}  % MRR ENG
        & 33.97\textsubscript{$\pm$0.27} % MRR AVG
        & -3.61 % Delta
        % & 33.42\textsubscript{$\pm$0.38}  % nDCG@5 ENG
        % & 32.03\textsubscript{$\pm$0.30} % nDCG@5 AVG
        % & -4.16 % Delta
        & 40.09\textsubscript{$\pm$0.35}  % nDCG@10 ENG
        & 38.59\textsubscript{$\pm$0.29}  % nDCG@10 AVG
        & -3.74 % Delta
        \\
        \hdashline

        \manner\textsubscript{\xlmrlarge}
        & 310
        & 562
        & 60.76\textsubscript{$\pm$9.32} % AUC ENG
        & 59.96\textsubscript{$\pm$8.62} % AUC AVG
        & -1.31 % Delta
        & 35.54\textsubscript{$\pm$0.59}  % MRR ENG
        & 34.02\textsubscript{$\pm$0.65}  % MRR AVG
        & -4.27 % Delta
        % & 33.58\textsubscript{$\pm$0.59}  % nDCG@5 ENG
        % & 32.18\textsubscript{$\pm$0.67} % nDCG@5 AVG
        % & -4.18 % Delta
        & 39.84\textsubscript{$\pm$0.47}  % nDCG@10 ENG
        & 38.41\textsubscript{$\pm$0.76}  % nDCG@10 AVG
        & -3.59 % Delta
        \\

        \manner\textsubscript{\labse}
        & 415
        & 472
        & 68.31\textsubscript{$\pm$1.26} % AUC ENG
        & 66.89\textsubscript{$\pm$1.21} % AUC AVG
        & -2.09 % Delta
        & 36.52\textsubscript{$\pm$0.65}  % MRR ENG
        & 35.24\textsubscript{$\pm$0.52}  % MRR AVG
        & -3.50 % Delta
        % & 34.91\textsubscript{$\pm$0.70}  % nDCG@5 ENG
        % & 33.59\textsubscript{$\pm$0.53} % nDCG@5 AVG
        % & -3.78 % Delta
        & 41.11\textsubscript{$\pm$0.65}  % nDCG@10 ENG
        & 39.81\textsubscript{$\pm$0.48}  % nDCG@10 AVG
        & -3.16 % Delta
        \\

        \manner\textsubscript{\nase}
        & 415
        & 472
        & 67.22\textsubscript{$\pm$0.40} % AUC ENG
        & 65.29\textsubscript{$\pm$0.68} % AUC AVG
        & -2.87 % Delta
        & 35.41\textsubscript{$\pm$0.61}  % MRR ENG
        & 33.86\textsubscript{$\pm$0.44}  % MRR AVG
        & -4.36 % Delta
        % & 33.49\textsubscript{$\pm$0.38}  % nDCG@5 ENG
        % & 32.12\textsubscript{$\pm$0.39} % nDCG@5 AVG
        % & -4.09 % Delta
        & 39.88\textsubscript{$\pm$0.41}  % nDCG@10 ENG
        & 38.44\textsubscript{$\pm$0.36}  % nDCG@10 AVG
        & -3.59 % Delta
        \\  
        
    \bottomrule
    \end{tabular}%
    }
    
    \vspace{-0.5em}
\end{table*}

%% file: tables/zsxlt_rec_frozen_lm.tex
\begin{table*}
    \centering

    \caption{\zsxlt{} recommendation performance with a frozen news encoder (LM/SE). For each model, we report its size in terms of millions of trainable and total parameters, and performance (i) on the English MIND (\texttt{ENG}), (ii) averaged across all 14 target languages of \xmind{} (\texttt{AVG}), and (iii) the relative percentage difference between average \zsxlt{} and \texttt{ENG} performance ($\%\Delta$). Reported performance is averaged over three runs. Subscripts denote standard deviation.}
    \label{tab:zsxlt_rec_frozen_lm}
    
    \resizebox{\textwidth}{!}{%
    \begin{tabular}{lllcgacgacga}
        \toprule
        \multirow{2}{*}{\textbf{Model}} & \multicolumn{2}{c}{\textbf{\# Parameters (M)}}
        & \multicolumn{3}{c}{\textbf{AUC}} & \multicolumn{3}{c}{\textbf{MRR}} 
        % & \multicolumn{3}{c}{\textbf{nDCG@5}} 
        & \multicolumn{3}{c}{\textbf{nDCG@10}} 
        \\ \cmidrule(lr){2-3} \cmidrule(lr){4-6} \cmidrule(lr){7-9} \cmidrule(lr){10-12} 
        & Trainable & Total
        & ENG & AVG & $\%\Delta$ 
        & ENG & AVG & $\%\Delta$ 
        % & ENG & AVG & $\%\Delta$ 
        & ENG & AVG & $\%\Delta$ \\ \hline
        
        \lfrecscl\textsubscript{\xlmrlarge}
        & 0
        & 559
        & 50.00\textsubscript{$\pm$0.00} % AUC ENG
        & 50.00\textsubscript{$\pm$0.00} % AUC AVG
        & -0.00 % Delta
        & 24.82\textsubscript{$\pm$0.00}  % MRR ENG
        & 24.64\textsubscript{$\pm$0.00}  % MRR AVG
        & -0.72 % Delta
        % & 22.17\textsubscript{$\pm$0.00}  % nDCG@5 ENG
        % & 22.27\textsubscript{$\pm$0.00} % nDCG@5 AVG
        % & +0.75% Delta
        & 28.44\textsubscript{$\pm$0.00}  % nDCG@10 ENG
        & 28.71\textsubscript{$\pm$0.00}  % nDCG@10 AVG
        & +0.92 % Delta
        \\

        \lfrecscl\textsubscript{\labse}
        & 0
        & 471
        & 50.47\textsubscript{$\pm$0.00} % AUC ENG
        & 50.98\textsubscript{$\pm$0.00}  % AUC AVG
        & +1.00 % Delta
        & 30.96\textsubscript{$\pm$0.00}  % MRR ENG
        & 28.99\textsubscript{$\pm$0.00} % MRR AVG
        & -6.35 % Delta
        % & 28.62\textsubscript{$\pm$0.00}  % nDCG@5 ENG
        % & 26.64\textsubscript{$\pm$0.00} % nDCG@5 AVG
        % & -6.92 % Delta
        & 34.83\textsubscript{$\pm$0.00}  % nDCG@10 ENG
        & 32.95\textsubscript{$\pm$0.00}  % nDCG@10 AVG
        & -5.41 % Delta
        \\

        \lfrecscl\textsubscript{\nase}
        & 0
        & 471
        & 64.19\textsubscript{$\pm$0.00} % AUC ENG
        & 63.79\textsubscript{$\pm$0.00} % AUC AVG
        & -0.62 % Delta
        & 34.74\textsubscript{$\pm$0.00}  % MRR ENG
        & 33.96\textsubscript{$\pm$0.00}  % MRR AVG
        & -2.26 % Delta
        % & 33.02\textsubscript{$\pm$0.00}  % nDCG@5 ENG
        % & 32.14\textsubscript{$\pm$0.00} % nDCG@5 AVG
        % & -2.66 % Delta
        & 39.01\textsubscript{$\pm$0.00} % nDCG@10 ENG
        & 38.23\textsubscript{$\pm$0.00}  % nDCG@10 AVG
        & -2.00 % Delta
        \\ \hdashline

        \naml\textsubscript{\xlmrlarge}
        & 4.9
        & 564
        & 50.01\textsubscript{$\pm$0.01} % AUC ENG
        & 50.01\textsubscript{$\pm$0.01} % AUC AVG
        & 0.00 % Delta
        & 26.75\textsubscript{$\pm$1.11}  % MRR ENG
        & 26.59\textsubscript{$\pm$0.92} % MRR AVG
        & -0.62 % Delta
        % & 24.58\textsubscript{$\pm$1.18}  % nDCG@5 ENG
        % & 24.29\textsubscript{$\pm$0.92} % nDCG@5 AVG
        % & -1.21 % Delta
        & 31.06\textsubscript{$\pm$1.26}  % nDCG@10 ENG
        & 30.93\textsubscript{$\pm$0.89}  % nDCG@10 AVG
        & -0.42 % Delta
        \\

        \naml\textsubscript{\labse}
        & 0.156
        & 471
        & 61.17\textsubscript{$\pm$0.01} % AUC ENG
        & 58.78\textsubscript{$\pm$0.70} % AUC AVG
        & -3.91 % Delta
        & 34.57\textsubscript{$\pm$0.09}  % MRR ENG
        & 33.86\textsubscript{$\pm$0.32} % MRR AVG
        & -2.05 % Delta
        % & 32.90\textsubscript{$\pm$0.02}  % nDCG@5 ENG
        % & 32.08\textsubscript{$\pm$0.32} % nDCG@5 AVG
        % & -2.49 % Delta
        & 39.21\textsubscript{$\pm$0.05}  % nDCG@10 ENG
        & 38.19\textsubscript{$\pm$0.35}  % nDCG@10 AVG
        & -2.60 % Delta
        \\

        \naml\textsubscript{\nase}
        & 0.156
        & 471
        & 63.30\textsubscript{$\pm$0.14} % AUC ENG
        & 61.05\textsubscript{$\pm$0.33} % AUC AVG
        & -3.55 % Delta
        & 33.89\textsubscript{$\pm$0.16} % MRR ENG
        & 32.73\textsubscript{$\pm$0.21}  % MRR AVG
        & -3.43 % Delta
        % & 31.98\textsubscript{$\pm$0.12}  % nDCG@5 ENG
        % & 31.04\textsubscript{$\pm$0.21} % nDCG@5 AVG
        % & -2.94 % Delta
        & 38.47\textsubscript{$\pm$0.10} % nDCG@10 ENG
        & 37.47\textsubscript{$\pm$0.18}  % nDCG@10 AVG
        & -2.60  % Delta
        \\
        \hdashline

        \mins\textsubscript{\xlmrlarge}
        & 8.8
        & 568
        & 49.93\textsubscript{$\pm$0.28} % AUC ENG
        & 50.30\textsubscript{$\pm$0.22} % AUC AVG
        & +0.75 % Delta
        & 29.77\textsubscript{$\pm$0.42}  % MRR ENG
        & 28.13\textsubscript{$\pm$1.49}  % MRR AVG
        & -5.51 % Delta
        % & 27.66\textsubscript{$\pm$0.51}  % nDCG@5 ENG
        % & 26.33\textsubscript{$\pm$1.59} % nDCG@5 AVG
        % & -4.79 % Delta
        & 33.96\textsubscript{$\pm$0.52}  % nDCG@10 ENG
        & 32.74\textsubscript{$\pm$1.47}  % nDCG@10 AVG
        & -3.57 % Delta
        \\

        \mins\textsubscript{\labse}
        & 2.7
        & 473
        & 61.25\textsubscript{$\pm$1.19} % AUC ENG
        & 60.37\textsubscript{$\pm$1.38} % AUC AVG
        & -1.43 % Delta
        & 32.69\textsubscript{$\pm$0.35}  % MRR ENG
        & 32.00\textsubscript{$\pm$0.51}  % MRR AVG
        & -2.09 % Delta
        % & 30.99\textsubscript{$\pm$0.35}  % nDCG@5 ENG
        % & 30.21\textsubscript{$\pm$0.55} % nDCG@5 AVG
        % & -2.51 % Delta
        & 37.37\textsubscript{$\pm$0.39}  % nDCG@10 ENG
        & 36.50\textsubscript{$\pm$0.55}  % nDCG@10 AVG
        & -2.34 % Delta
        \\

        \mins\textsubscript{\nase}
        & 2.7
        & 473
        & 60.19\textsubscript{$\pm$0.64} % AUC ENG
        & 58.62\textsubscript{$\pm$0.81} % AUC AVG
        & -2.61 % Delta
        & 32.23\textsubscript{$\pm$0.54}  % MRR ENG
        & 31.65\textsubscript{$\pm$0.68}  % MRR AVG
        & -1.80 % Delta
        % & 30.52\textsubscript{$\pm$0.62}  % nDCG@5 ENG
        % & 29.91\textsubscript{$\pm$0.75} % nDCG@5 AVG
        % & -2.00 % Delta
        & 37.08\textsubscript{$\pm$0.55}  % nDCG@10 ENG
        & 36.44\textsubscript{$\pm$0.70}  % nDCG@10 AVG
        & -1.73 % Delta
        \\
        \hdashline

        \caum\textsubscript{\xlmrlarge}
        & 8.2
        & 559
        & 55.14\textsubscript{$\pm$0.61} % AUC ENG
        & 55.00\textsubscript{$\pm$0.67} % AUC AVG
        & -0.27 % Delta
        & 28.55\textsubscript{$\pm$0.71}  % MRR ENG
        & 28.40\textsubscript{$\pm$0.76}  % MRR AVG
        & -0.53 % Delta
        % & 26.61\textsubscript{$\pm$0.69}  % nDCG@5 ENG
        % & 26.48\textsubscript{$\pm$0.77} % nDCG@5 AVG
        % & -0.52 % Delta
        & 33.26\textsubscript{$\pm$0.59}  % nDCG@10 ENG
        & 33.14\textsubscript{$\pm$0.69}  % nDCG@10 AVG
        & -0.36 % Delta
        \\

        \caum\textsubscript{\labse}
        & 3.7
        & 474
        & 62.82\textsubscript{$\pm$0.77} % AUC ENG
        & 61.35\textsubscript{$\pm$0.74} % AUC AVG
        & -2.34 % Delta
        & 33.79\textsubscript{$\pm$0.59}  % MRR ENG
        & 33.00\textsubscript{$\pm$0.48}  % MRR AVG
        & -2.33 % Delta
        % & 31.92\textsubscript{$\pm$0.65}  % nDCG@5 ENG
        % & 31.16\textsubscript{$\pm$0.54} % nDCG@5 AVG
        % & -2.40 % Delta
        & 38.40\textsubscript{$\pm$0.56}  % nDCG@10 ENG
        & 37.61\textsubscript{$\pm$0.46}  % nDCG@10 AVG
        & -2.05 % Delta
        \\

        \caum\textsubscript{\nase}
        & 3.7
        & 474
        & 64.40\textsubscript{$\pm$0.61} % AUC ENG
        & 62.69\textsubscript{$\pm$0.64} % AUC AVG
        & -2.65 % Delta
        & 34.42\textsubscript{$\pm$0.44}  % MRR ENG
        & 33.49\textsubscript{$\pm$0.37}  % MRR AVG
        & -2.72 % Delta
        % & 32.73\textsubscript{$\pm$0.40}  % nDCG@5 ENG
        % & 31.63\textsubscript{$\pm$0.34} % nDCG@5 AVG
        % & -3.36 % Delta
        & 39.15\textsubscript{$\pm$0.43}  % nDCG@10 ENG
        & 37.98\textsubscript{$\pm$0.34}  % nDCG@10 AVG
        & -2.98 % Delta
        \\
        \hdashline

        \manner\textsubscript{\xlmrlarge}
        & 2.1
        & 562
        & 62.11\textsubscript{$\pm$1.17}  % AUC ENG
        & 50.80\textsubscript{$\pm$0.30} % AUC AVG
        & -18.21 % Delta
        & 30.88\textsubscript{$\pm$1.42}  % MRR ENG
        & 23.86\textsubscript{$\pm$0.27}  % MRR AVG
        & -22.73 % Delta
        % & 29.26\textsubscript{$\pm$1.31}  % nDCG@5 ENG
        % & 21.22\textsubscript{$\pm$0.29} % nDCG@5 AVG
        % & 27.46% Delta
        & 35.56\textsubscript{$\pm$1.18}  % nDCG@10 ENG
        & 27.78\textsubscript{$\pm$0.25}  % nDCG@10 AVG
        & -21.88 % Delta
        \\

        \manner\textsubscript{\labse}
        & 1.7
        & 472
        & 64.54\textsubscript{$\pm$0.97} % AUC ENG
        & 61.97\textsubscript{$\pm$0.85} % AUC AVG
        & -3.98 % Delta
        & 33.96\textsubscript{$\pm$0.87} % MRR ENG
        & 31.45\textsubscript{$\pm$0.76} % MRR AVG
        & -7.37 % Delta
        % & 32.25\textsubscript{$\pm$0.82}  % nDCG@5 ENG
        % & 29.62\textsubscript{$\pm$0.71} % nDCG@5 AVG
        % & -8.15  % Delta
        & 38.51\textsubscript{$\pm$0.77} % nDCG@10 ENG
        & 35.82\textsubscript{$\pm$0.67} % nDCG@10 AVG
        & -6.98 % Delta
        \\

        \manner\textsubscript{\nase}
        & 1.7
        & 472
        & 65.51\textsubscript{$\pm$1.07} % AUC ENG
        & 64.14\textsubscript{$\pm$0.78} % AUC AVG
        & -2.08 % Delta
        & 34.68\textsubscript{$\pm$0.89}  % MRR ENG
        & 33.45\textsubscript{$\pm$0.72}  % MRR AVG
        & -3.54  % Delta
        % & 32.94\textsubscript{$\pm$0.92} % nDCG@5 ENG
        % & 31.78\textsubscript{$\pm$0.71} % nDCG@5 AVG
        % & -3.53 % Delta
        & 39.13\textsubscript{$\pm$0.77}  % nDCG@10 ENG
        & 37.94\textsubscript{$\pm$0.62}  % nDCG@10 AVG
        & -3.05 % Delta
        \\

    \bottomrule
    \end{tabular}
    }
    
    \vspace{-0.5em}
\end{table*}

%% file: sections/conclusion.tex
\section{Conclusion}
\label{sec:conclusion}
Current neural news recommenders based on multilingual language models (i) suffer from substantial performance losses in \zsxlt{} recommendation, and (ii) usually require expensive fine-tuning of the language model used as the news encoder backbone: such fine-tuning is often infeasible, e.g., in cold-start setups without click-behavior data. In this work, we proposed \nase{}, a news-adapted sentence encoder obtained through domain specialization of a pretrained multilingual sentence encoder. To this end, we compiled and leveraged two multilingual news-specific corpora, \polynews{} and \polynewsparallel{}. Our findings question the effectiveness of supervised fine-tuning for news recommendation. As an efficient solution, we proposed a simple and strong baseline based on frozen \nase{} embeddings and late click behavior fusion that achieves state-of-the-art performance in \zsxlt{} in true cold start and few-shot news recommendation.

%% file: sections/acknowledgements.tex
\subsubsection{\ackname}
\label{sec:acknowledgements}

The authors acknowledge support from the state of Baden-Württemberg through bwHPC and the German Research Foundation (DFG) through grant INST 35/1597-1 FUGG.

%% file: sections/appendix.tex
\section{Appendix}
\label{sec:appendix_polynews}

\subsection{Language Characteristics}
\label{sec:appendix_polynews_lang}

Table \ref{tab:polynews_language_characteristics} lists the languages included in the \polynews{} dataset, summarizing the following information, according to the \#BenderRule:
\begin{itemize}
    \item \textbf{Code:} We denote each language using a BCP 47 tag sequence that combines a base subtag indicating the three-letter ISO 693-3 code with the ISO 15924 script subtag. We use the script code to differentiate between languages collected in multiple scripts.
    \item \textbf{Language:} In case of multiple denominations, we use the language name from Ethnologue \cite{lewis2009ethnologue}, cross-referenced against other linguistic resources, namely Glottolog \cite{hammarstrom2021glottolog} and World Atlas of Structures (WALS) \cite{wals}.
    \item \textbf{Script:} We provide the English name of the script.
    \item \textbf{Macro-area:} We indicate the macro-area as per WALS \cite{wals}.
    \item \textbf{Family and Subgrouping:} We specify the language family and subgrouping from WALS \cite{wals} and Glottolog \cite{hammarstrom2021glottolog}.
    \item \textbf{Total Speakers:} We report the total number of speakers of the language, considering both L1-level (first-language) and L2-level (second-language) speakers, according to Ethnologue.\footnote{We use the statistics available in May 2024 at \href{https://www.ethnologue.com/}{https://www.ethnologue.com/}.}
    \item \textbf{\labse{} support:} We indicate whether the language was included in the pretraining corpora of \labse{} \cite{feng-etal-2022-language}.
    \item \textbf{Average byte length:} We report the average number of bytes per text for each language.
    \item \textbf{Average character length:} We report the average number of characters per text for each language.
\end{itemize}

\input{tables/polynews_language_characteristics}

\subsection{Parallel Corpora Statistics}
\label{sec:appendix_polynewsparallel_stats}

\begin{figure*}[]
    \centering
    \includegraphics[width=\textwidth]{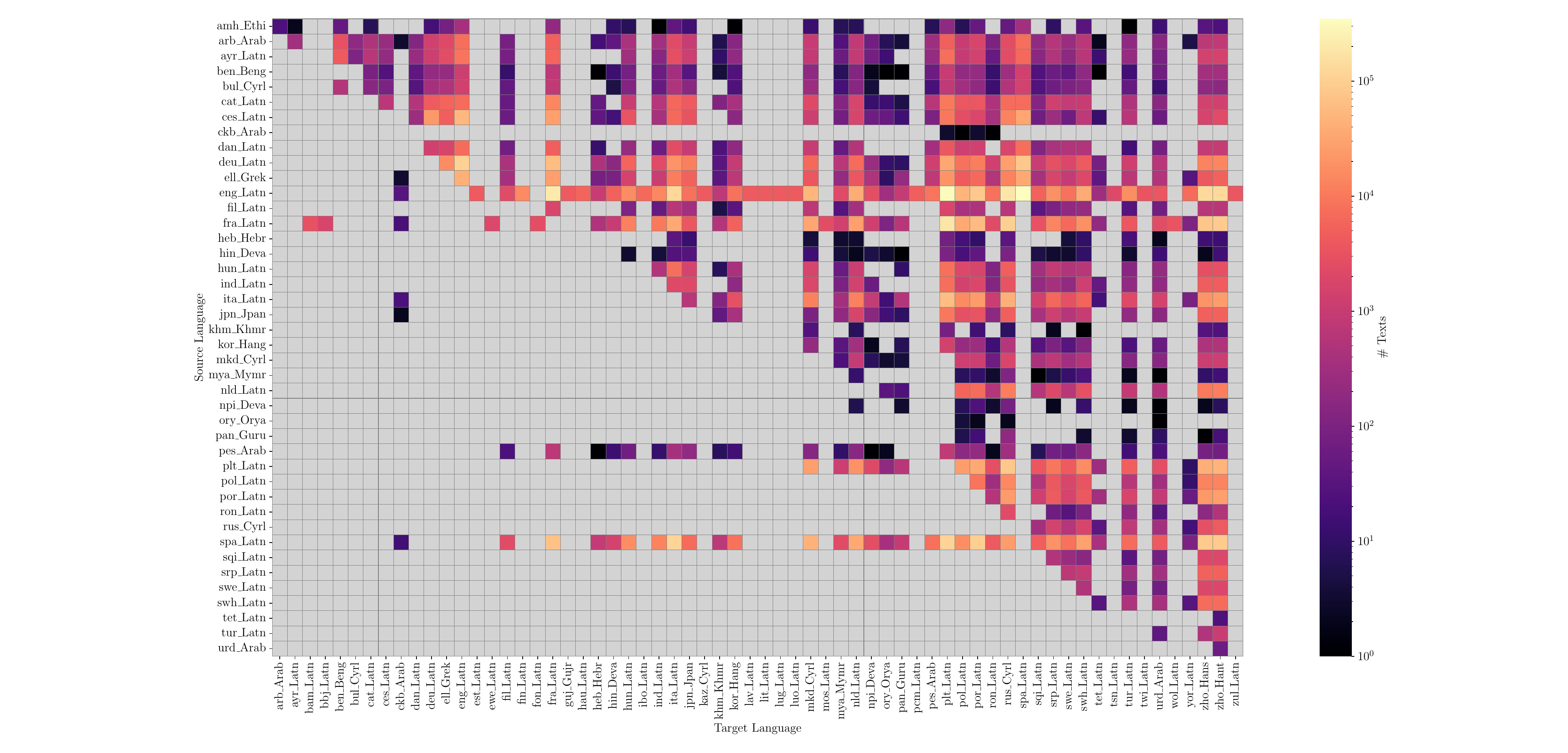}
    \caption{Distribution of texts across the 833 language pairs in \polynewsparallel{}. The gray cells indicate that no texts exist for the given language pair.}
    \label{fig:polynewsparallel_language_distrib}
    \vspace{-0.5em}
\end{figure*}

Fig. \ref{fig:polynewsparallel_language_distrib} illustrates the number of texts for each language pair available in the \polynewsparallel{} dataset. Note that \polynewsparallel{} consists of only 64 out of the 77 languages listed in Table \ref{tab:polynews_language_characteristics}.

%% file: tables/polynews_language_characteristics.tex
\begin{table*}[t!]
\centering
\resizebox{\textwidth}{!}{%  
\begin{tabular}{lllllllcll}
    \toprule
    \textbf{Code} & \textbf{Language} & \textbf{Script} & \textbf{Macro-area} & \textbf{Family} & \textbf{Subgrouping} & \textbf{\# Speakers (M)} & \textbf{\labse} & \textbf{Avg. \# bytes}  & \textbf{Avg. \# chars.}\\
    \midrule
    
     amh\_Ethi & \textbf{Amharic} & Ethiopic & Africa & Afro-Asiatic & Semitic & 59.69 & \cmark{} & 190.42 & 80.10  \\
     
     arb\_Arab & \textbf{Modern Standard Arabic} & Arabic & Eurasia & Afro-Asiatic & Semitic & 332.46 & \cmark{} & 225.86 &  130.60  \\
     
     ayr\_Latn & \textbf{Central Aymara} & Latin & South America & Aymaran & Central Southern Aymara & 1.46 & \xmark{} & 122.33 &  118.48  \\
     
     bam\_Latn & \textbf{Bambara} & Latin & Africa & Mande & Western Mande & 14.19 & \xmark{}  & 135.94 & 122.92  \\
     
     bbj\_Latn & \textbf{Ghomálá’} & Latin & Africa & Niger-Congo & Atlantic-Congo & 0.35 & \xmark{}  & 86.77 & 64.87  \\
     
     ben\_Beng & \textbf{Bengali} & Bengali & Eurasia & Indo-European & Indo-Aryan & 278.18 & \cmark{}  & 300.72 & 128.01  \\
     
     bos\_Latn & \textbf{Bosnian} & Latin & Eurasia & Indo-European & Balto-Slavic & 2.68 & \cmark{}  & 52.08 & 50.76  \\
     
     bul\_Cyrl & \textbf{Bulgarian} & Cyrillic & Eurasia & Indo-European & Balto-Slavic & 7.88 & \cmark{}  & 256.64 &  148.65 \\
     
     cat\_Latn & \textbf{Catalan} & Latin & Eurasia & Indo-European & Italic & 9.29 & \cmark{}  & 132.07 &  129.25 \\
     
     ces\_Latn & \textbf{Czech} & Latin & Eurasia & Indo-European & Balto-Slavic & 12.32 & \cmark{} & 127.28 &  114.53  \\

     \rowcolor{Gray}
     ckb\_Arab & \textbf{Central Kurdish} & Arabic & Eurasia & Indo-European & Indo-Iranian & 5.35 & \cmark{}  & 350.93 &  185.93 \\
     
     dan\_Latn & \textbf{Danish} & Latin & Eurasia & Indo-European & Germanic & 5.81 & \cmark{}  & 133.51 &  130.71 \\
     
     deu\_Latn & \textbf{German} & Latin & Eurasia & Indo-European & Germanic & 133.91 & \cmark{}  & 136.53 &  134.47 \\
     
     ell\_Grek & \textbf{Greek} & Greek & Eurasia & Indo-European & Greek & 13.23 & \cmark{}  & 273.36 & 159.73  \\
    
     eng\_Latn & \textbf{English} & Latin & Eurasia & Indo-European & Germanic & 1,515.23 & \cmark{}  & 123.89 & 123.28  \\
     
     est\_Latn & \textbf{Estonian} & Latin & Eurasia & Uralic & Finnic & 1.09 (L1 only) & \cmark{}  & 116.43 & 113.24  \\
     
     ewe\_Latn & \textbf{Éwé} & Latin & Africa & Niger-Congo & Atlantic-Congo & 5.53 & \xmark{}  &  78.44 & 72.39  \\ 
     
     fil\_Latn & \textbf{Filipino} & Latin & Papunesia & Austronesian & Malayo-Polynesian & -- & \xmark{}  & 135.54 & 135.24  \\
     
     fin\_Latn & \textbf{Finnish} & Latin & Eurasia & Uralic & Finnic & 5.61 & \cmark{}  & 117.52 & 112.76  \\
     
     fon\_Latn & \textbf{Fon} & Latin & Africa & Niger-Congo & Atlantic-Congo & 2.28 & \xmark{} & 179.91 &  136.43  \\

     fra\_Latn & \textbf{French} & Latin & Eurasia & Indo-European & Italic & 311.58 & \cmark{}  & 143.82  & 139.25  \\
     
     guj\_Gujr & \textbf{Gujarati} & Gujarati & Eurasia & Indo-European & Indo-Iranian & 62.67 & \cmark{}  & 301.16 & 118.81  \\
     
     guw\_Latn & \textbf{Gun} & Latin & Africa & Niger-Congo & Atlantic-Congo & 1.54 & \xmark{}  & 93.61  & 76.95  \\
     
     hau\_Latn & \textbf{Hausa} & Latin & Africa & Afro-Asiatic & Chadic & 88.24 & \cmark{}  & 138.36  & 137.40  \\
     
     heb\_Hebr & \textbf{Hebrew} & Hebrew & Eurasia & Afro-Asiatic & Semitic & 9.36 & \cmark{}  & 132.72 & 77.33  \\
     
     hin\_Deva & \textbf{Hindi} & Devanagari & Eurasia & Indo-European & Indo-Iranian & 608.85 & \cmark{}  & 321.32 &  129.44 \\
     
     hun\_Latn & \textbf{Hungarian} & Latin & Eurasia & Uralic & -- & 12.43 & \cmark{}  & 141.26 &  128.87 \\
     
     ibo\_Latn &\textbf{Igbo}  & Latin & Eurasia & Niger-Congo & Attlantic-Congo & 30.91 & \cmark{}  & 123.64 &  104.73 \\
     
     ind\_Latn & \textbf{Indonesian} & Latin & Papunesia & Austronesian & Malayo-Polynesian & 199.00 & \cmark{} & 131.97 &  131.56  \\
     
     ita\_Latn & \textbf{Italian} & Latin & Eurasia & Indo-European & Italic & 66.79 & \cmark{}  & 138.79 & 137.64  \\
     
     jpn\_Jpan & \textbf{Japanese} & Japanese & Eurasia & Japonic & Japanesic  & 123.47 & \cmark{}  & 160.54 &  64.63 \\
     
     kaz\_Cyrl & \textbf{Kazakh} & Cyrillic & Eurasia & Altaic & Turkic & 16.58 & \cmark{}  & 279.77 & 155.47  \\
     
     khm\_Khmr & \textbf{Khmer} & Khmer & Eurasia & 	Austro-Asiatic & Khmeric & 17.60 & \cmark{}  & 368.65 &  139.12 \\
     
     kor\_Hang & \textbf{Korean} & Hangul & Eurasia & Koreanic & Korean & 81.13 & \cmark{}  & 156.69 & 72.81  \\
     
     lav\_Latn & \textbf{Latvian} & Latin & Eurasia & Indo-European & Baltic & 1.76 & \cmark{}  & 135.63 & 124.39  \\
     
     lin\_Latn & \textbf{Lingala} & Latin & Africa & Niger-Congo & Atlantic-Congo & 40.54 & \xmark{}  & 66.14 & 65.62  \\   
     
     lit\_Latn & \textbf{Lithuanian} & Latin & Eurasia & Indo-European & Balto-Slavic & 2.78 & \cmark{}  & 135.97 & 126.83  \\
     
     lug\_Latn & \textbf{Ganda} & Latin & Africa & Niger-Congo & Atlantic-Congo & 11.14 & \xmark{}  & 111.21 & 110.90  \\
     
     luo\_Latn & \textbf{Luo} & Latin & Africa & Nilotic & Western Nilotic & 5.26 & \xmark{}  & 158.18 & 157.76  \\
     
     mkd\_Cyrl & \textbf{Macedonian} & Cyrillic & Eurasia & Indo-European & Balto-Slavic & 1.65 & \cmark{}  & 261.49 & 150.68  \\
     
     mos\_Latn & \textbf{Mossi} & Latin & Africa & Niger-Congo & Atlantic-Congo & 11.87 & \xmark{}  & 140.87 & 127.62  \\
     
     mya\_Mymr & \textbf{Burmese} & Myanmar & Eurasia & Sino-Tibetan & Tibeto-Burman & 43.17 & \cmark{}  & 383.44 & 148.70  \\
     
     nld\_Latn & \textbf{Dutch} & Latin & Eurasia & Indo-European & Germanic & 25.31 & \cmark{}  & 124.46 & 123.98  \\
     
     nor\_Latn & \textbf{Norwegian} & Latin & Eurasia & Indo-European & Germanic & 5.42 & \cmark{}  & 44.14 &  43.20 \\
     
     npi\_Deva & \textbf{Nepali} & Devanagari & Eurasia &  Indo-European & Indo-Iranian & 31.98 & \cmark{}  & 441.96 & 178.46  \\
     
     orm\_Latn & \textbf{Oromo} & Latin & Africa & Afro-Asiatic & Cushitic & 45.53 (L1 only) & \xmark{}  & 79.86 & 79.57  \\
     
     \rowcolor{Gray}
     ory\_Orya & \textbf{Odia} & Oriya & Eurasia & Indo-European & Indo-Iranian & 39.76 & \cmark{}  &  268.32 & 107.68  \\
     
     pan\_Guru & \textbf{Eastern Panjabi} & Gurmukhi & Eurasia & Indo-European & Indo-Iranian & 31.17 & \cmark{}  & 685.89 & 275.91  \\
     
     pcm\_Latn & \textbf{Nigerian Pidgin} & Latin & Africa & Creole & -- & 120.65 & \xmark{}  & 118.72 & 118.69   \\
     
     pes\_Arab & \textbf{Western Persian} & Arabic & Eurasia & Indo-European & Indo-Iranian & 78.83 & \cmark{}  & 218.93 &  128.05 \\ 
     
     plt\_Latn & \textbf{Malagasy} & Latin & Africa & Austronesia & Malayo-Polinesian & 7.55 & \cmark{}  & 152.57 & 151.77  \\
     
     pol\_Latn & \textbf{Polish} & Latin & Eurasia & Indo-European & Balto-Slavic & 41.80 & \cmark{}  & 120.85 & 114.56  \\
     
     por\_Latn & \textbf{Portuguese} & Latin & Eurasia & Indo-European & Italic & 263.84 & \cmark{} & 124.35 & 120.73   \\
     
     ron\_Latn & \textbf{Romanian} & Latin & Eurasia & Indo-European & Italic & 25.23 & \cmark{} & 129.89 & 123.63   \\
     
     run\_Latn & \textbf{Rundi} & Latin & Africa & Niger-Congo & Atlantic-Congo & 13.19 & \xmark{}  & 75.73 &  75.20 \\
     
     rus\_Cyrl & \textbf{Russian} & Cyrillic & Eurasia & Indo-European & Balto-Slavic & 255.39 & \cmark{}  & 151.75 & 88.92  \\
     
     sna\_Latn & \textbf{Shona} & Latin & Africa & Niger-Congo & Atlantic-Congo & 10.88 & \cmark{}  & 67.94 & 67.77  \\
     
     som\_Latn & \textbf{Somali} & Latin & Africa & Afro-Asiatic & Cushitic & 23.88 & \cmark{}  & 73.09 & 72.98  \\
     
     spa\_Latn & \textbf{Spanish} & Latin & Eurasia & Indo-European & Italic & 559.52 & \cmark{}  &136.71 & 134.00  \\

     sqi\_Latn & \textbf{Albanian} & Latin & Eurasia & Indo-European & Albanian & 6.39 & \cmark{}  & 128.83  & 119.94  \\

     srp\_Cyrl & \textbf{Serbian} & Cyrillic & Eurasia & Indo-European & Balto-Slavic & 10.12 & \cmark{}  & 88.89 & 51.97  \\
     
     srp\_Latn & \textbf{Serbian} & Latin &  Eurasia & Indo-European & Balto-Slavic & 10.12 & \cmark{}  & 78.22 & 76.17  \\

     swe\_Latn & \textbf{Swedish}  & Latin & Eurasia & Indo-European & Germanic & 13.25 & \cmark{}  &  112.77 & 108.45  \\
     
     swh\_Latn & \textbf{Swahili} & Latin & Africa & Niger-Congo & Atlantic-Congo & 86.52 & \cmark{}  & 115.32 & 115.00  \\
     
     tam\_Taml & \textbf{Tamil} & Tamil & Eurasia & Dravidian & Southern Dravidian & 86.72 & \cmark{}  & 170.92 & 64.02  \\
     
     tet\_Latn & \textbf{Tetun} & Latin & Africa & Austronesian & Malayo-Polynesian & 0.49 & \xmark{}  & 130.19 &  128.88 \\

    \rowcolor{Gray}
     tha\_Thai & \textbf{Thai} & Thai & Eurasia & Kra-Dai & Kam-Tai & 61.10 & \cmark{}  & 144.90 & 56.57   \\

     \rowcolor{Gray}
     tir\_Ethi & \textbf{Tigrinya} & Ethiopic & Africa & Afro-Asiatic & Semitic & 9.87 & \xmark{}  & 116.47 & 47.10  \\
   
     tsn\_Latn & \textbf{Tswana} & Latin & Africa & Niger-Congo & Atlantic-Congo & 13.75 & \xmark{} & 142.02 & 141.69   \\
     
     tur\_Latn & \textbf{Turkish} & Latin & Eurasia & Altaic & Turkic & 90.09 & \cmark{}  & 122.87 &  112.45 \\
     
     twi\_Latn & \textbf{Twi} & Latin & Africa & Niger-Congo & Atlantic-Congo & 0.024 & \xmark{}  & 119.62 & 109.66  \\
     
     ukr\_Cyrl & \textbf{Ukrainian} & Cyrillic & Eurasia & Indo-European & Balto-Slavic & 38.92 & \cmark{}  & 107.62 &  60.99 \\
     
     urd\_Arab & \textbf{Urdu} & Arabic & Eurasia & Indo-European & Indo-Iranian & 169.70 & \cmark{}  & 193.15 &  115.04 \\ 

      wol\_Latn & \textbf{Wolof} & Latin & Africa & Niger-Congo & Atlantic-Congo & 22.65 & \cmark{}  & 128.05 & 122.11  \\
     
     xho\_Latn & \textbf{Xhosa} & Latin & Africa & Niger-Congo & Atlantic-Congo & 19.22 & \cmark{}  & 47.79 &  47.37 \\
     
     yor\_Latn & \textbf{Yorùbá} & Latin & Africa & Niger-Congo & Atlantic-Congo & 47.19 & \cmark{} & 174.46 & 124.68   \\

      zho\_Hans & \textbf{Chinese} & Han (Simplified) & Eurasia & Sino-Tibetan & Sinitic &1,352.67 (L1 only)  & \cmark{}  & 78.06 &  36.56 \\
     
     zho\_Hant & \textbf{Chinese} & Han (Traditional) & Eurasia & Sino-Tibetan & Sinitic & 1,352.67 (L1 only) & \xmark{}  & 76.86 & 36.64  \\
     
     zul\_Latn & \textbf{Zulu} & Latin & Africa & Niger-Congo & Atlantic-Congo & 27.80 & \cmark{}  & 165.13 & 164.64  \\
     
     \bottomrule

\end{tabular}%
}
\caption{The 77 languages of \polynews. We display the language \textit{Code} (ISO 693-3), language name, \textit{Script}, \textit{Macro-area}, \textit{Family} and \textit{Genus}, and report the total number of L1-level and L2-level speakers of the language according to Ethnologue \cite{lewis2009ethnologue}. The eighth column indicates whether the language was included in the pretraining corpora of \labse{} \cite{feng-etal-2022-language}. The last two columns specify the average byte and character lengths of the texts for each language. The languages highlighted in gray were not included in the adaptive pretraining of \nase.}
\label{tab:polynews_language_characteristics}
\vspace{-0.5em}
\end{table*}